\documentclass[twocolumn,showpacs,prl,aps,floats,superscriptaddress]{revtex4}

\usepackage{epsfig}
\usepackage{amsmath}
\usepackage{amssymb}

\begin{document}

\title{System-size dependence of strangeness production \\ in nucleus-nucleus
collisions at $\sqrt{s_{NN}}=17.3$ GeV}

\affiliation{NIKHEF, Amsterdam, Netherlands.}
\affiliation{Department of Physics, University of Athens, Athens,
Greece.} \affiliation{Comenius University, Bratislava, Slovakia.}
\affiliation{KFKI Research Institute for Particle and Nuclear
Physics, Budapest, Hungary.} \affiliation{MIT, Cambridge, MA,
USA.} \affiliation{Institute of Nuclear Physics, Cracow, Poland.}
\affiliation{Gesellschaft f\"{u}r Schwerionenforschung (GSI),
Darmstadt, Germany.} \affiliation{Joint Institute for Nuclear
Research, Dubna, Russia.} \affiliation{Fachbereich Physik der
Universit\"{a}t, Frankfurt, Germany.} \affiliation{CERN, Geneva,
Switzerland.} \affiliation{University of Houston, Houston, TX,
USA.} \affiliation{\'{S}wietokrzyska Academy, Kielce, Poland.}
\affiliation{Fachbereich Physik der Universit\"{a}t, Marburg,
Germany.} \affiliation{Max-Planck-Institut f\"{u}r Physik, Munich,
Germany.} \affiliation{Institute of Particle and Nuclear Physics,
Charles University, Prague, Czech Republic.}
\affiliation{Department of Physics, Pusan National University,
Pusan, Republic of Korea.} \affiliation{Nuclear Physics
Laboratory, University of Washington, Seattle, WA, USA.}
\affiliation{Atomic Physics Department, Sofia University St.
Kliment Ohridski, Sofia, Bulgaria.} \affiliation{Institute for
Nuclear Studies, Warsaw, Poland.} \affiliation{Institute for
Experimental Physics, University of Warsaw, Warsaw, Poland.}
\affiliation{Rudjer Boskovic Institute, Zagreb, Croatia.}

\author{C.~Alt}
    \affiliation{Fachbereich Physik der Universit\"{a}t, Frankfurt, Germany.}
\author{T.~Anticic}
        \affiliation{Rudjer Boskovic Institute, Zagreb, Croatia.}
\author{B.~Baatar}
    \affiliation{Joint Institute for Nuclear Research, Dubna, Russia.}
\author{D.~Barna}
        \affiliation{KFKI Research Institute for Particle and Nuclear Physics, Budapest, Hungary.}
\author{J.~Bartke}
        \affiliation{Institute of Nuclear Physics, Cracow, Poland.}
\author{L.~Betev}
        \affiliation{CERN, Geneva, Switzerland.}
        \affiliation{Fachbereich Physik der Universit\"{a}t, Frankfurt, Germany.}
\author{H.~Bia{\l}\-kowska}
        \affiliation{Institute for Nuclear Studies, Warsaw, Poland.}
\author{A.~Billmeier}
        \affiliation{Fachbereich Physik der Universit\"{a}t, Frankfurt, Germany.}
\author{C.~Blume}
        \affiliation{Gesellschaft f\"{u}r Schwerionenforschung (GSI), Darmstadt, Germany.}
        \affiliation{Fachbereich Physik der Universit\"{a}t, Frankfurt, Germany.}
\author{B.~Boimska}
        \affiliation{Institute for Nuclear Studies, Warsaw, Poland.}
% ...

\collaboration{NA49 Collaboration} \noaffiliation

\date{\today}

\begin{abstract}

Emission of $\pi^{\pm}$, K$^{\pm}$, $\phi$ and $\Lambda$ was
measured in near-central C+C and Si+Si collisions at 158~$A$GeV
beam energy. Together with earlier data for p+p, S+S and Pb+Pb,
the system-size dependence of relative strangeness production in
nucleus-nucleus collisions is obtained. Its fast rise and the
saturation observed at about 60 participating nucleons can be
understood as onset of the formation of coherent systems of
increasing size.

\end{abstract}

\pacs{25.75.-q}

\maketitle

Most recent experimental studies of particle production in
relativistic nucleus-nucleus collisions have focussed on heavy
reaction partners because of the expectation that the transition
to a deconfined state of strongly interacting matter \cite{qcd} is
more likely to occur in large collision systems. However, earlier
experiments with lighter beams at the CERN SPS accelerator had
demonstrated that already in $^{32}$S+S collisions \cite{na35_ss}
relative strangeness production  $-$ which has been discussed as a
possible indicator for such a transition \cite{rafelskimueller}
$-$ is significantly enhanced compared to p+p collisions and, as
found later, is close to the situation in $^{208}$Pb+Pb
\cite{na49_energy,na49_lambda}.

In order to understand the origin of this enhancement a systematic
study of its dependence on the size of the collision system is
required. In this Letter we discuss systems with similar geometry,
i.\,e.~near-central and approximately symmetric $A+A$ collisions
between nuclei of varying mass number $A$.  New data on the
production of $\pi^{\pm}$, K$^{\pm}$, $\phi$-mesons and
$\Lambda$-baryons will be presented for the systems C+C and Si+Si.
They were obtained with the NA49 spectrometer at the CERN SPS at
158~$A$GeV beam energy, corresponding to a center-of-mass energy
of $\sqrt{s_{NN}} = 17.3$ GeV. The results will be discussed
together with available data for p+p and Pb+Pb at 158 $A$GeV and
S+S at 200 $A$GeV beam energy. It will be shown that strangeness
enhancement develops fast with increasing size of the collision
system reaching the level of Pb+Pb already at about 60
participating nucleons. We argue that this result can be
understood assuming the formation of coherent systems of
increasing size.


NA49 is a fixed-target experiment at CERN \cite{na49_nim} using
external SPS beams. For the present study a fragmented Pb beam of
158 $A$GeV was used \cite{na49_nim}. The fragments were selected
by magnetic rigidity ($Z/A= 0.5$) and by specific energy loss in
transmission detectors in the beam line. The "C-beam" as defined
by the online and offline trigger was a mixture of ions with $Z$=6
and 7 (intensity ratio 69:31) in the case of the meson
measurements, and pure $Z$=6 ions for the $\Lambda$. As "Si-beam"
we used ions with $Z$=13,14, and 15 (35:41:24). 3~mm and 10~mm
thick carbon targets (2.4\% and 7.9\% interaction length) were
chosen, and a 5~mm thick Si target (4.4\%). The veto calorimeter
placed 20~m downstream of the target accepts nearly all beam
particles, projectile fragments and spectator neutrons and protons
\cite{na49_nim}. By setting an upper threshold for the energy
deposit, the ($15.3\pm 2.4$)\% most central C+C and ($12.2\pm
1.8$)\% most central Si+Si interactions were selected \cite{diss}.
After all cuts typically 45k events were analyzed for the mesons
and 200k for the $\Lambda$. Simulations based on the VENUS model
of nucleus-nucleus interactions \cite{venus} allow to calculate
impact parameter range and mean number of wounded nucleons
\cite{bialas}: $b<2.0$~fm for C+C and $b<2.6$~fm for Si+Si, for
$\langle N_{\rm wound}\rangle$ see table \ref{yields}. In $A+A$
reactions some nucleons in the collision may participate only
through secondary cascading processes. In this paper, we do not
include such nucleons in counting the number of wounded nucleons
$N_{\rm wound}$, but we do include them in the number of
participating nucleons $N_{\rm part}$, which is also provided by
VENUS. These definitions differ from those assumed for Glauber
calculations. For the $N_{\rm part}$ values given in table
\ref{yields} the numbers from VENUS are averaged with those
derived from the measured energy of the projectile spectators in
the veto calorimeter.

Charged particles produced in beam-target interactions were
measured in the NA49 large-acceptance hadron spectrometer
consisting of magnets and time projection chambers (TPCs), two of
them immediately behind the target and inside the magnetic field,
two additional ones further downstream on both sides of the beam
(for details see \cite{na49_nim}). By tracking and measurement of
the specific energy loss ($dE/dx$) in the TPCs the momentum and
the identity of the particles were determined. In the
(lab)momentum range $4<p<50$~GeV/$c$ and with a relative $dE/dx$
resolution ranging from about (3-6)\% depending on track length
and momentum \cite{na49_nim} pions, kaons and protons could be
resolved on a statistical basis by decomposing the $dE/dx$ spectra
in $(p,p_{t})$ bins which are then transformed to $(y,p_{t})$ or
$(y,m_{t})$ bins ($p_{t}$ = transverse momentum, $m_{t}$ =
transverse mass, $y$ = c.m.\,rapidity). Within the geometrical
acceptance of the spectrometer which covers most of the forward
hemisphere in the center-of-mass system of the collision pions and
kaons were selected in a range of the azimuthal angle (about 28\%
of 2$\pi$) where small background and a flat acceptance is
ensured. Acceptance and corrections for in-flight decay of kaons
(about 2\% losses in addition) were determined by GEANT
\cite{geant} simulations. The contribution of pions from weak
decays of hadrons (mainly K$^{0}_{S}$) is subtracted using events
from the VENUS model (about 5\%). Care is taken that the VENUS
multiplicities agree with data, otherwise they are scaled
accordingly. The $\phi$-mesons were measured via their strong
decay into K$^{+}$K$^{-}$ pairs (branching ratio 49.1\%). Yields
were derived in $y$- or $p_{t}$-bins from the corresponding peak
in the K$^{+}$K$^{-}$ invariant-mass spectra after subtracting the
combinatorial background; the overall acceptance in the forward
hemisphere amounts to about 45\%. The technique is described in
\cite{na49_phi}; for more results also on the other mesons see
\cite{diss}. $\Lambda$-hyperons decay weakly via $\Lambda \to
\text{p}\pi^{-}$ (branching ratio 63.9\%). They were identified on
the basis of their typical V0 decay topology and the p$\pi^{-}$
invariant mass; for the method see \cite{na49_lambda}. The
$p_{t}$-integrated combination of acceptance and detection
efficiency ranges from 8\% in the most backward $y$-bin to about
57\% at midrapidity. Since all $\Sigma^{0}$ decay
electromagnetically to $\Lambda$-baryons, $\Lambda$ denotes the
sum of both. For Si+Si an additional correction of (5$\pm$3)\% for
the reduced $\Lambda$ reconstruction efficiency was applied due to
the higher track density in these reactions. GEANT simulations
showed that, varying slightly with $p_{t}$ and $y$, 90\% of the
$\Lambda$-baryons from weak decays of $\Xi^{-}$ and $\Xi^{0}$
contribute to the measured $\Lambda$-yield. Using yield estimates
from statistical model fits \cite{becattini_privat} a feeddown
correction of (9$\pm$3)\% and (10$\pm$3)\% was applied for
$\Lambda$ from C+C and Si+Si, respectively.

\begin{figure}[]
\epsfig{file=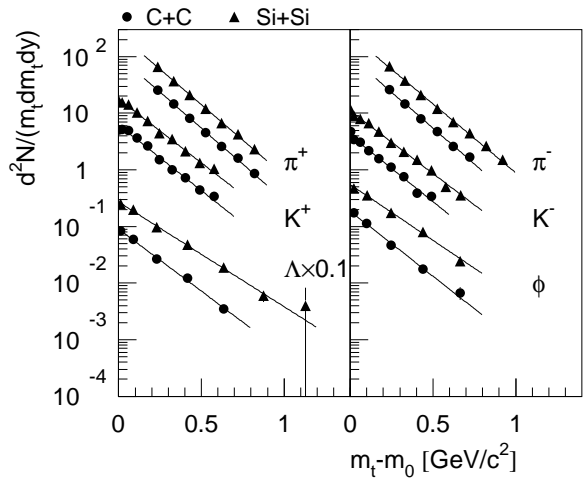,width=8cm} \vspace{-0.2cm}
\caption{\label{mt} Transverse mass distributions for $\pi$ and K
in $0.1<y<0.33$, for $\phi$ averaged over $0<y<1.8$, and for
$\Lambda$ in $|y|<0.4$ measured in C+C ({\large$\bullet$}) and
Si+Si ($\blacktriangle$) collisions. Only statistical errors are
shown.}
\end{figure}
\begin{figure}
\epsfig{bbllx=0,bblly=0,bburx=310,bbury=410,file=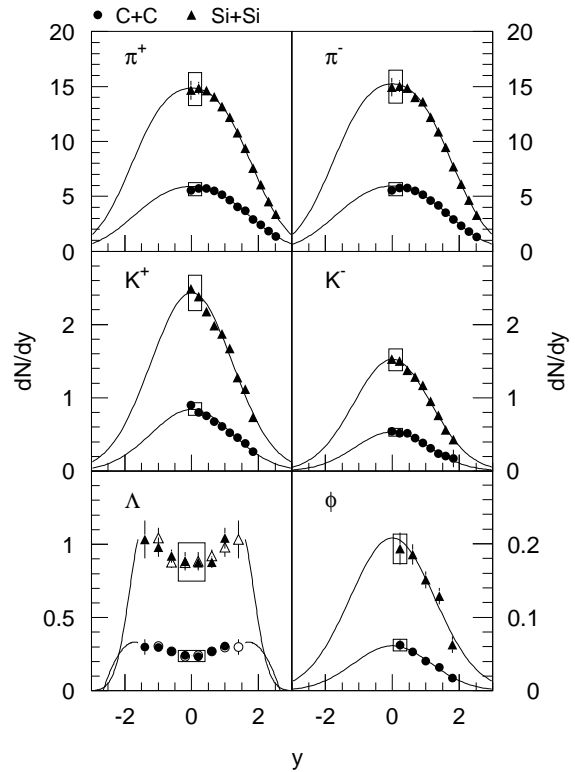,width=8cm}
\vspace{-0.4cm} \caption{\label{y} Rapidity distributions for C+C
({\large$\bullet$}) and Si+Si ($\blacktriangle$) collisions. Open
symbols are reflected at midrapidity. Error bars are statistical,
systematic errors at midrapidity are indicated by the rectangular
boxes. The curves represent fits to the data as described in the
text.}
\end{figure}

For all particles under study yields were obtained in $(y,p_{t})$
bins. The transverse distributions, those close to midrapidity
being shown in fig.~\ref{mt},
 were fitted with a thermal ansatz
$d^{2}N/(m_{t}dm_{t}dy) \propto \exp(-m_{t}/T)$. The fit was used
to extrapolate to $p_{t}$-regions not covered by the measurement,
and the yield $dN/dy$ was obtained by summing the measured bins
and this extrapolation. The latter correction was of the order of
a few percent typically. An exception were pions at midrapidity
($-0.13<y<0.33$) for which the correction factors became as large
as 2 to 3 due to the limit $p<4$ GeV/$c$ for particle
identification. In this case, spectral shapes from the NA35 S+S
\cite{na35_pi} measurements were used, this way taking the known
low-$p_{t}$ pion enhancement into account. In order to estimate
the systematic error in this phase space region different analysis
strategies were investigated: When extrapolating the pion yields
with a straight exponential with an inverse slope $T$ as given in
table \ref{spectra}, yields would be lower by (6-9)\%. An $h^{-}$
analysis as performed in \cite{na49_energy} with corrections for
$K^{-}$ and $\bar{p}$ gives (5-9)\% higher yields. This
uncertainty mainly determines the systematic error on the
midrapidity and total pion yields. The pion rapidity distributions
$dN/dy$ (fig.~\ref{y})
 were approximated by a superposition of two Gaussians
displaced symmetrically around mid-rapidity by a shift $\pm
y_{0}$. For kaons and the $\phi$-meson a single Gaussian was used.
Integration gives the total average yield $\langle N\rangle$ of
each particle. To obtain the yields of $\Lambda$-baryons their
$y$-distribution was fitted by a shape adopted from p+p and S+S
measurements. The mean of both approximations was taken for C+C;
for Si+Si only the shape from S+S was used (lines in
fig.~\ref{y}). For the $\phi$-meson with its much lower abundance
the $p_{t}$-distribution could only be obtained when averaged over
a wider rapidity range, $0<y<1.8$. The values for $dN/dy$ were
determined in $y$-slices using the $p_{t}$-range $0<p_{t}<1.5$
GeV/$c$ and the averaged $T$-parameter for extrapolation
(fig.~\ref{y}).

Parameters for the kinetic distributions of all studied particles
are given in table \ref{spectra},
\begin{table}[t]
 \begin{ruledtabular}
  \begin{tabular}{ c c c c c }
    C+C & $T$ [MeV] & $dN/dy$ & $\sigma_{y}$ & $y_{0}$ \\
    \hline
    $\pi^{+}$ & 171 $\pm$ 10 & 5.6 $\pm$ 0.6 & 1.12 $\pm$ 0.09 & 0.87 $\pm$ 0.07\\
    $\pi^{-}$ & 171 $\pm$ 10 & 5.7 $\pm$ 0.6 & 1.11 $\pm$ 0.09 & 0.85 $\pm$ 0.07 \\
    $\text{K}^{+}$ & 188 $\pm$ 10 & 0.85 $\pm$ 0.085 & 1.23 $\pm$ 0.08 & \\
    $\text{K}^{-}$ & 185 $\pm$ 10 & 0.53 $\pm$ 0.053 & 1.10 $\pm$ 0.09 & \\
    $\phi$ & 189 $\pm$ 28 & 0.062 $\pm$ 0.008 & 1.16 $\pm$ 0.10 & \\
    $\Lambda$ & 199 $\pm$ 15 & 0.24 $\pm$ 0.038 & & \\
    \hline
    Si+Si & $T$ [MeV] & $dN/dy$ & $\sigma_{y}$ & $y_{0}$ \\
    \hline
    $\pi^{+}$ & 173 $\pm$ 10 & 14.8 $\pm$ 1.5 & 1.04 $\pm$ 0.08 & 0.91 $\pm$ 0.07 \\
    $\pi^{-}$ & 178 $\pm$ 10 & 15.0 $\pm$ 1.5 & 1.05 $\pm$ 0.08 & 0.89 $\pm$ 0.07 \\
    $\text{K}^{+}$ & 192 $\pm$ 10 & 2.4 $\pm$ 0.24 & 1.24 $\pm$ 0.08 &  \\
    $\text{K}^{-}$ & 196 $\pm$ 10 & 1.5 $\pm$ 0.15 & 1.15 $\pm$ 0.07 & \\
    $\phi$ & 220 $\pm$ 28 & 0.19 $\pm$ 0.02 & 1.27 $\pm$ 0.10 & \\
    $\Lambda$ & 235 $\pm$ 16 & 0.88 $\pm$ 0.13 & & \\
  \end{tabular}
 \end{ruledtabular}
\vspace{-0.2cm}
 \caption
 {\label{spectra} Parameters of the measured kinetic distributions. For pions and kaons $T$ holds for
 $0.1<y<0.33$, $dN/dy$ for $-0.13<y<0.33$; for $\Lambda$
 both values are given for $|y|<0.4$.
 For the $\phi$-meson $T$ is determined in
 $0<y<1.8$, $dN/dy$ in $0<y<0.4$. Only the dominating errors are given,
 i.e.\,statistical errors for the $\phi$ and systematic errors
 for the other hadrons. Statistical errors of the latter are less than
 half the systematic error; for the $\phi$ the systematic error is
 either the same ($dN/dy$) or half of the statistical error.}
\end{table}
 their yields in table
\ref{yields}.
\begin{table}[t]
 \begin{ruledtabular}
  \begin{tabular}{ c c c }
    & C+C & Si+Si \\
    \hline
    $\langle N_{\rm wound} \rangle$ & 14 $\pm$ 2 & 37 $\pm$ 3 \\
    $\langle N_{\rm part} \rangle$ & 16.3 $\pm$ 1 & 41.4 $\pm$ 2 \\
    \hline
    $\langle\pi^{+}\rangle$ & 22.4 $\pm$ 0.3 $\pm$ 1.6 & 56.6 $\pm$ 0.7 $\pm$ 4 \\
    $\langle\pi^{-}\rangle$ & 22.2 $\pm$ 0.3 $\pm$ 1.6 & 57.6 $\pm$ 0.6 $\pm$ 4\\
    $\langle\text{K}^{+}\rangle$ & 2.54 $\pm$ 0.03 $\pm$ 0.25 & 7.44 $\pm$ 0.08 $\pm$ 0.74\\
    $\langle\text{K}^{-}\rangle$ &  1.49 $\pm$ 0.05 $\pm$ 0.15 & 4.42 $\pm$ 0.04 $\pm$ 0.44\\
    $\langle\phi\rangle$ &  0.18 $\pm$ 0.01 $\pm$ 0.02 & 0.66 $\pm$ 0.03 $\pm$ 0.08\\
    $\langle\Lambda\rangle$ &  1.32 $\pm$ 0.05 $\pm$ 0.32 & 3.88 $\pm$ 0.16 $\pm$ 0.56  \\
  \end{tabular}
 \end{ruledtabular}
\vspace{-0.2cm}
 \caption
 {\label{yields} Particle yields for C+C (15.3\% most central collisions) and
 Si+Si (12.2\%) at $\sqrt{s_{NN}}=17.3$ GeV. The
 first error is statistical, the second systematic.}
\end{table}

\begin{figure}
\epsfig{bbllx=0,bblly=0,bburx=330,bbury=285,file=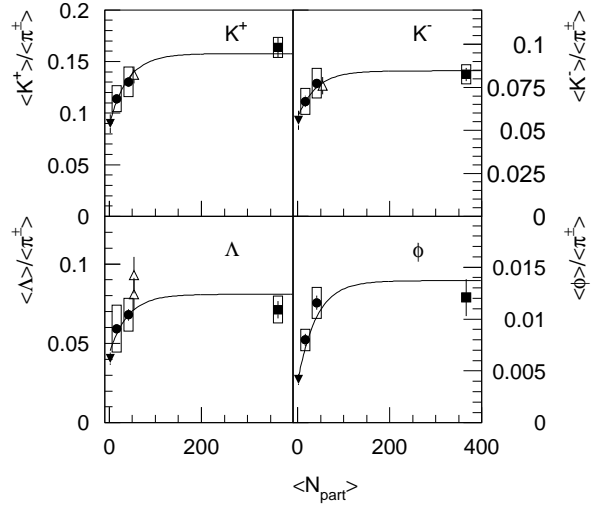,width=8cm}
\vspace{-0.2cm} \caption{\label{ratios} Experimental ratios of
$\langle\text{K}^{+}\rangle$, $\langle\text{K}^{-}\rangle$,
$\langle\phi\rangle$, and $\langle\Lambda\rangle$ to
$\langle\pi^{\pm}\rangle$ plotted as a function of system size
({\footnotesize$\blacktriangledown$} p+p, {\large$\bullet$} C+C
and Si+Si, {\footnotesize$\triangle$} S+S, {\tiny$\blacksquare$}
Pb+Pb). Statistical errors are shown as error bars, systematic
errors if available as rectangular boxes. The curves are shown to
guide the eye and represent a functional form $a -
b\cdot\exp(-\langle N_{\rm part}\rangle/40)$. At $\langle N_{\rm
part}\rangle = 60$ they rise to about 80\% of the difference of
the ratios between $N_{\rm part}= 2$ and 400.}
\end{figure}

The strange-particle yields were normalized to the mean number of
charged pions per event, $\langle\pi^{\pm}\rangle =
(\langle\pi^{+}\rangle+\langle\pi^{-}\rangle)/2$. In this way they
represent the relative contribution of strange quarks in the final
state. The results are shown in fig.~\ref{ratios}
together with data for minimum-bias p+p
\cite{rossi,na49_phi,na49_lambda} and Pb+Pb interactions (5\% most
central, $\Lambda$-baryons 10\% most central)
\cite{na49_energy,na49_phi,na49_lambda} taken at the same energy.
The $\Lambda$-yield is scaled by the ratio of the number of
wounded nucleons to account for the difference in the centrality
selection, and the correction for feeddown from $\Xi$-baryons is
applied \cite{na49_lambda}. Also plotted are the data for S+S (2\%
most central) obtained at the slightly higher energy of
$\sqrt{s_{NN}}=19.4$ GeV \cite{na35_ss}. $\Lambda$-yields for S+S
were scaled down by 10\% to account for feeddown from
$\Xi$-baryons assuming the same $\Xi/\Lambda$ ratio as in Si+Si
and using the information on the magnitude of the feeddown
contribution given in \cite{na35_ss}.

Using the mean number of nucleons participating in the collision
($\langle N_{\rm part}\rangle$) as a measure of system size the
resulting dependence of the relative strangeness production
(fig.~\ref{ratios}) is characterized by a fast rise at small
$\langle N_{\rm part}\rangle$ values, reaching the level of Pb+Pb
interactions at $\langle N_{\rm part}\rangle \sim 60$. Note that
the same shape of the enhancement curve holds for all measured
strangeness-carrying particles including the $\phi$-meson with
only hidden strangeness.

The system-size dependence of relative strangeness production
indicates basic changes in the reaction mechanism due to the
transition from an isolated N+N interaction to the situation in
$A+A$ where nucleons undergo several subsequent collisions within
short time intervals and where many of these collision sequences
occur close in space. Microscopic transport models such as UrQMD
\cite{urqmd} calculate position and time of each collision in the
center-of-mass system of the reaction. This allows to extract a
collision density per unit space and time interval averaged over
the interpenetration phase of the colliding nuclei. This collision
density is found to increase with the size of the colliding
nuclei; for central Si+Si reactions about 4-5 interactions per
fm$^{3}$ and fm/$c$ are reached \cite{qm02_nantes, diss}. It is
reasonable to assume that neither the subsequent collisions nor
the decay of the created objects proceed independently. We propose
that the reaction volume should thus become quantum-mechanically
coherent. Apparently, these conditions influence the global
strangeness production rate.

Statistical models for hadron production
(e.g.~\cite{pbm,becattini}) are very successful in reproducing the
particle ratios in p+p as well as in $A+A$ collisions. In these
models the implementation of strangeness conservation results in a
strong volume dependence of strangeness production
\cite{rafelski,hagedorn,redlich}. It is natural to invoke this
effect, known as canonical strangeness suppression, as a
qualitative explanation for the data shown in fig.~\ref{ratios}.
Quantitatively, however, discrepancies remain. In the model of
Tounsi and Redlich \cite{redlich}, based on a hadronic scenario,
as well as in the model of Rafelski and Danos \cite{rafelski},
based on both, a partonic and a hadronic scenario, the strangeness
saturation point (as defined in the caption of fig.~\ref{ratios})
for $|S|=1$ particles is close to $N_{\rm part}= 10$, in contrast
to the experimental value of $N_{\rm part}=60$. One can reconcile
theory with experiment assuming that only parts of the reaction
volume form the statistical ensembles used in the models. This
implies a modification of the linear relationship between volume
and number of participants which was assumed so far.


To summarize, the data obtained for near-central symmetric
nucleus-nucleus collisions with increasing number of participating
nucleons show a steep rise of the relative strangeness content in
the produced hadrons reaching a saturation value at around
$\langle N_{\rm part}\rangle \sim 60$. It is argued that this
behavior is correlated with the formation of extended systems of
dense and excited matter which collectively decay to the observed
hadrons. Within the statistical model for hadron production this
would qualitatively explain the increase of relative strangeness
production, because the constraints due to strangeness
conservation are diminishing with increasing size of the
hadronizing systems. Strangeness saturation thus occurs when the
grand-canonical limit can be applied
\cite{rafelski,hagedorn,redlich,stock}. The experiment shows that
for singly strange particles and the $\phi$-meson at
$\sqrt{s_{NN}}=17.3$~GeV this happens already in central
collisions between nuclei with $A$ about 30.

\begin{acknowledgments}
This work was supported by the US Department of Energy Grant
DE-FG03-97ER41020/A000, the Bundesministerium fur Bildung und
Forschung, Germany, the Polish State Committee for Scientific
Research (2 P03B 130 23, SPB/CERN/P-03/Dz 446/2002-2004, 2 P03B
04123), the Hungarian Scientific Research Foundation (T032648,
T032293, T043514), the Hungarian National Science Foundation,
OTKA, (F034707), the Polish-German Foundation, and the Korea
Research Foundation Grant (KRF-2003-041-C00088).
\end{acknowledgments}


\end{document}